\documentstyle[epsf,amssymb,12pt]{article}
\def\.{\!\cdot\!}
\def\:{\cdots}
\def\[{\left[}
\def\]{\right]}
\def\({\left(}
\def\){\right)}
\def\bk#1{\langle#1\rangle}
\def\bi{\begin{itemize}}

\def\ei{\end{itemize}}
\def\be{\begin{eqnarray}}
\def\ee{\end{eqnarray}}
\def\bn{\begin{enumerate}}
\def\en{\end{enumerate}}
\def\h{{1\over 2}}

\def\nn{\nonumber}

\def\r2{\sqrt{2}}

\def\x{\times}
\def\labels#1{\label{#1}}
\def\ket#1{|#1\rangle}
\def\bra#1{\langle#1|}
\def\eq#1{(\ref{#1})}

\def\t{\tau}
\def\D{\Delta}
\def\d{\delta}
\def\l{\lambda}
\def\Z{\bf Z}

\begin{document}
\begin{center}
{\bf{Neutrino Oscillations via the Bulk}}\\
\vspace{.5cm}
C.S. Lam$^1$ and J.N. Ng$^2$\\
\vspace{.2cm}
 $^1$\sl{Department of Physics, McGill University, Montreal\\
\sl {Email: Lam@physics.mcgill.ca}\\
and\\
 $^2$\sl {Theory Group, TRIUMF, Vancouver}\\
\sl Email: Misery@triumf.ca}
\end{center}
\vspace{.3 cm}

\begin{abstract}
We investigate the possibility that the large mixing of neutrinos
is induced by their large coupling to a five-dimensional bulk neutrino.
In the strong coupling limit the model is exactly soluble. It gives
rise to an oscillation amplitude whose squared-mass difference is
independent of the channel, thus making it impossible to explain
both the solar and the atmospheric neutrino oscillations simultaneously.
\end{abstract}


\section{Introduction}
The idea of using an extra dimension to explain physics goes back to
Kaluza and Klein \cite{KK}. Spurred by the discovery of D-branes in
string theory \cite{POL}, this idea was revived in recent years by
introducing branes \cite{AK} with many variations to the theme.
We will adopt here the simplest one, where a single three-brane is
embedded in an extra flat dimension. All Standard-Model (SM) particles are
confined to our 3+1 dimensional world, the three-brane. Gravity,
right-handed neutrinos, and other particles without SM
quantum numbers are allowed to live in the 4+1 dimensional bulk. 
This modification thus introduced to gravity has been tested experimentally
\cite{WASH}. Its consequence on astrophysics and cosmology is extensively
investigated \cite{COS}. In this paper we will discuss an application
of this picture to neutrino physics \cite{KEITH,MO,NEU}.

If solar and atmospheric neutrino deficits are attributed to neutrino
oscillations, then neutrinos must be massive. Direct measurement puts
an upper bound of about 2 eV on $\nu_e$ \cite{MAINZ}. If it is
a Majorana particle, the failure to
detect neutrinoless double $\beta$ decay further limits a 
combination of the lightest neutrinos and mixings to have a mass less 
than 0.26 eV \cite{HM}. In any case
neutrino mass is much smaller than other fermion masses and this 
mystery should be understood. Moreover, the amount of mixing needed
to explain the atmospheric neutrino deficit is large \cite{SK}, 
whereas the mixings of quarks are small.
Presently these two differences between neutrinos and other fermions
are not clearly understood.

In the seesaw mechanism, the smallness of neutrino mass
is explained by the high energy scale of the right-handed neutrinos.
The smallness can also be attributed to the large size of the extra
dimension \cite{KEITH,MO,NEU}, and it would be nice if the extra dimension 
can also
explain the large neutrino mixing. It is this latter possibility
that we carefully examine in the present paper.

To this end we study the simplest model, 
which consists of one bulk neutrino and three active Majorana brane
neutrinos, $\nu_e,\nu_\mu$, and $\nu_\t$. 
Since mixings between quarks are small, to simplify matters we
shall assume no direct mixings between the active
neutrinos in this model. All mixings between
them are induced solely through their mass couplings to the bulk neutrino.
The bulk neutrino is a Dirac fermion in 4+1 dimensions, 
and a SM singlet. All in all,
the theory contains six parameters, the
Majorana masses $m_i$ for the three active neutrinos, and their 
mass couplings $d_i$ to the bulk neutrino. For equal couplings this model 
was discussed by Dienes et al in \cite{KEITH}. [Other similar models were also
studied \cite{MO}.]  Without Majorana masses and sometimes
with only one generation, this model has also been
studied by several other groups \cite{MO}.
 The  mass couplings $d_i$
modify  neutrino masses and
induce a mixing between them.  If the large mixings 
seen in experiments are to be explained by this simple
model, the  mass coupling strengths must be large.  
We recall that $d_i=y_iv/\sqrt{2 \pi M_* R}$
where $y_i$ is the Yukawa coupling between the $i$-th brane neutrino 
and the bulk neutrino,
$v$ is the usual vacuum expectation value of the SM Higgs field, 
$M_*$ is the fundamental 5D scale and
$R$ is the compactification radius. Hence, the strong coupling case  will 
arise from large brane- bulk
coupling and usual perturbative treatments assuming $d_i \ll 1$ will not 
apply. 
It is this strong-coupling scenario which we find most interesting and will
be the main focus of this paper.

Numerical studies have been carried out in \cite{KEITH,MO} for specific
parameters. 
Our analysis is completely analytical and we find remarkably that 
there is actually an exact solution to the problem
in the strong-coupling limit that we are interested in. 
In that limit, we are able to give a complete 
discussion of the eigenvalues and eigenvectors, as well as exact
brane-to-brane transition amplitudes which we have not found 
in the literature.

In the absence of coupling, the mass
eigenvalues are integers 
 for the infinite number of sterile neutrinos composing
the bulk neutrino, and $m_i$ for the active neutrinos confined to the
brane. All masses are understood to be measured in units of the
inverse size of the extra dimension.
In the strong-coupling limit, sterile and active components
are thoroughly mixed up. The mass eigenvalues 
now consist of every half-integer,
plus two others, $\l_b$ and $\l_c$, determined by $m_i$ and the
fixed ratios of the $d_i$'s. More generally, if there are $f$ 
active neutrinos then there
are $f-1$ such other eigenvalues.
With this knowledge, 
the transition amplitudes from active
flavor $i$ to active flavor $j$ ($i,j=\nu_e,\nu_\mu,\nu_\t$)
can be computed exactly at any `time' $\t$. It contains  three
terms. The first comes from a sum of all transitions via 
half-integral eigenvalues. It is short lived, and it decays to zero
within a short `time' inversely proportional to the fourth power of
the coupling. The second and the third terms are transitions
mediated
by eigenvalues $\l_b$ and $\l_c$. 
After the transient period when
the first term dies away, we are left with an amplitude resembling the 
usual two-neutrino oscillating amplitude, but with a squared-mass
difference $\l_c^2-\l_b^2$ independent of $i$ and $j$. This 
independence makes it impossible to explain both the solar neutrino
deficit ($i=j=\nu_e$) and the atmospheric neutrino deficit ($i=j
=\nu_\mu$) simultaneously, for their measured
squared-mass differences are
at least two orders of magnitudes apart. 
There are actually also other reasons
why this model would not work in practice but they will be discussed
later in the main text.

The failure of this model to describe Nature is not due to the presence
of sterile neutrinos per se, for they are still allowed. Although
sterile neutrinos are disfavored in the solar and
atomospheric data of Super-Kamiokande \cite{SK}, solutions 
still exist in which
the solar oscillation occurs between $\nu_e$ and a sterile neutrino
\cite{BAHCALL}, and a substantial sterile component is also allowed
in atmospheric neutrino oscillations \cite{MARRONE}. If LSND
experiment is confirmed we do need a sterile component anyway. The
failure rather is due to the presence of an infinite number of them,
with mass eigenvalues at half integers.
It is the regular spacing of these eigenvalues at
large coupling that produces a destructive interference at large
$\t$, which eliminates the first term of the transition amplitude,
and kills the model. In this our conclusion is consistent
with earlier numerical investigations 
 and analytical studies for one active neutrino \cite{KEITH,MO}.
If the first term were oscillatory instead of decaying, which would
be the case if the infinite number of them
 were replaced by a single sterile
neutrino, then we would just get a regular three-neutrino oscillation
formula and a problem of this nature would not arise.

\section{Mass Matrix and Its Diagonalization}
We consider a model in which a three-brane is placed at an $S^1/Z_2$ 
orbifold fixed point. The size of the extra dimension is measured by
the radius $R$ of the circle $S^1$. The model
consists of three left-handed Majorana
neutrinos $\nu_e,\nu_\mu$, and $\nu_\t$,
confined to the brane, and a Dirac bulk neutrino $\Psi$ living in
the 4+1 dimensional world. The only interactions are mass terms,
and we shall express all masses here in units of $1/R$.
The bulk neutrino decomposes into an infinite number of 3+1 dimensional 
sterile neutrinos, with masses $n\in\Z$. The Majorana masses of
the brane neutrinos are denoted by $m_i$, and their couplings to
the bulk neutrino are denoted by the complex number $d_i$.
 Since phases are not currently detectable, we will assume $m_i$
and $d_i$ to be real. This practice is also adapted in numerical studies
\cite{KEITH,MO}.

The neutrino mass matrix is given by \cite{KEITH}
\be M=\pmatrix{m_1&0&0&d_1&d_1&d_1&d_1&d_1&\cdots\cr
               0&m_2&0&d_2&d_2&d_2&d_2&d_2&\cdots\cr
                 0&0&m_3&d_3&d_3&d_3&d_3&d_3&\cdots\cr
               d_1&d_2&d_3&0 &0 &0 &0 &0 &\cdots\cr
               d_1&d_2&d_3&0 &1 &0 &0 &0 &\cdots\cr
               d_1&d_2&d_3&0 &0 &-1 &0 &0 &\cdots\cr
               d_1&d_2&d_3&0 &0 &0 &2 &0 &\cdots\cr
               d_1&d_2&d_3&0 &0 &0 &0 &-2 &\cdots\cr
                 \cdots&&&&&&\cdots&\cdots\cr}.\labels{mm}\ee
Let $v=(v_1,v_2,\cdots)^T$ be its column eigenvector with
eigenvalue $\l$. Let $A=\sum_{i=4}^\infty v_i$, and 
$b=\sum_{i=1}^3d_iv_i$.
The eigenvalue equation $M\.v=\l v$ in component form is
\be
m_iv_i+d_i A&=&\l v_i, \qquad (i=1,2,3)\\
\labels{eigen123}
b+c_iv_i&=&\l v_i,\qquad (i\ge 4)\labels{eigen4}\ee
where $c_{2n}=2-n$ and $c_{2n+1}=n-1$ for $n\ge 2$. 
It follows from \eq{eigen4} that
\be
v_4&=&{b\over\l},\nn\\
v_{2n+3}&=&{b\over\l+n},\nn\\
v_{2n+4}&=&{b\over\l-n},\qquad (n\ge 1).
\labels{vlarge}\ee 

Hence
\be
A=\sum_{i=4}^\infty v_i=b\l\sum_{n=-\infty}^\infty{1\over\l^2-n^2}=
b{\pi\over\tan(\pi\l)}.\labels{a}\ee
Substituting this into \eq{eigen123}, it becomes
\be
M\,'v\equiv\pmatrix{m'_1+d_1^2&d_1d_2&d_1d_3\cr d_2d_1&m'_2+d_2^2&d_2d_3\cr
d_3d_1&d_3d_2&m'_3+d_3^2\cr}\.\pmatrix{v_1\cr v_2\cr v_3\cr}
=0,\labels{mpv}\ee
where
\be
m'_i=(m_i-\l)\tan(\pi\l)/\pi.\labels{mpi}\ee
The eigenvalue $\l$ is then determined by $\det[M']=0$. We shall
use the characteristic equation in the form
\be
{1\over\pi}\tan(\pi\l)=\sum_{i=1}^3{d_i^2\over\l-m_i}.\labels{charac}\ee

\subsection{Eigenvalues and Eigenvectors}
The eigenvalues can be determined in the following way. In the 
unit interval $(n-\h,
n+\h)$, $n\in\Z$,
the left-hand side of \eq{charac} increases from $-\infty$
to $+\infty$ as  $\l$ moves from the left edge to the right edge
of this interval.
On the other hand, the function on the right-hand side of \eq{charac}
is a monotonically 
decreasing function of $\l$ in each of the four intervals
$(-\infty,m_1),(m_1,m_2),(m_2,m_3),(m_3,+\infty)$, if
$m_1<m_2<m_3$. It decreases from 0 to $-\infty$ in the first interval,
from $+\infty$ to $-\infty$ in the second and third intervals, and
from $+\infty$ to 0 in the fourth interval. Hence if we divide the
real axis into intervals bounded by half integers (odd integers
divided by 2), as well as 
$m_1,m_2,m_3$, then there is one and only one solution of \eq{charac}
in each of these intervals.  See Fig.~1.

{}From \eq{mpv}, we can determine $(v_1,v_2,v_3)$ simply by taking the
cross product of any two rows of the matrix $M'$. Using \eq{mpi}
and \eq{charac}, the result (to within an arbitary normalization) can
be taken to be
\be
v_i={d_i\over\l-m_i},\qquad(i=1,2,3).\labels{v123}\ee
With this normalization, 
\be
b=\sum_{i=1}^3d_iv_i=\sum_{i=1}^3{d_i^2\over \l-m_i}={1\over\pi}\tan(\pi\l),
\labels{b}\ee
where \eq{charac} is used in the last expression.
The rest of the components, $v_i$ for $i\ge 4$, is determined by
\eq{vlarge}.
The norm of the vector $v$ is equal to
\be
B^2&=&v\.v=\sum_{i=1}^3\({d_i\over\l-m_i}\)^2
+{1\over\cos^2(\pi\l)}.\labels{norm}\ee
The second term in \eq{norm} is obtained from
\be
B_\perp^2&\equiv&\sum_{i=4}^\infty v_i^2=
b^2\sum_{n=-\infty}^\infty{\l^2+n^2\over(\l^2-n^2)^2}
=\({\pi b\over\sin(\pi\l)}\)^2\labels{normperp}\ee
by using \eq{b}.

\subsection{Unitary matrix}
It is more convenient to replace the flavor 
indices $i=1,2,3$ by $a_1,a_2,a_3$. Index $i=2n+3$ for $n\ge 1$ will be renamed
index $-n$, and index $i=2n+4$ for $n\ge 0$ will be renamed index $n$.
With this alteration, eqs.~\eq{v123} and \eq{vlarge} become
\be
v_{a_i}&=&{d_i\over \l-m_i}, \qquad (a_1,a_2,a_3)=(\nu_e,\nu_\mu,\nu_\t)\nn\\
v_n&=&{b\over\l-n},\qquad(n\in\Z).\labels{vnew}\ee
We shall refer to $a_i$ as the brane components, and 
$n$ as the bulk components.

The overlapping matrix between a normalized eigenvector $\ket{\l}$
and a flavor component $\bra{f}$ $(f=a_i,n)$ is then given by the unitary
matrix
\be
U_{f\lambda}&=&\bk{f|\lambda}={v_f(\l)\over B(\l)},\label{u}\ee
where $v_f(\l)$ is given by \eq{vnew} and $B(\l)$ is obtained from 
\eq{norm}.

\section{Neutrino Oscillations}
The probability for a brane component $a_j$ to turn into a brane component
$a_i$, after the neutrino of energy $E$ has travelled a distance $L$,
\be
{\cal P}_{ij}(\tau)&=&\bigg|{\cal A}_{ij}(\t)\bigg|^2,\labels{prob}\ee
is given in terms of the transition amplitude
\be{\cal A}_{ij}(\t)&=&\sum_\l 
U^*_{a_i\l}U_{a_j\l}e^{-i\l^2\tau},\labels{amp}\ee
where $\tau=L/2E$ for vacuum oscillations. 
The sum over eigenvalues $\l$ is an infinite
sum and can be evaluated only when the eigenvalues are explicitly known.

\section{Strong Coupling Limit}
When $d_i=0$, the neutrino mass matrix $M$ is diagonal. The eigenvalues
are $m_i$ for the brane states of active neutrinos,
 and $n\in\Z$ for the bulk states of sterile neutrinos. 
There is no mixing between any two flavor states.

When $d_i\not=0$, eigenvalues shift, and mixing between
flavor states are induced.  If $d_i$ are small, the shifts and mixings
can be calculated by perturbation theory. This however is not the 
interesting regime because it will never generate
the observed neutrino flavor mixings, which are large.

When $d_i$ are big, there is a chance that such large mixings can be 
induced. This is the regime we will study in the present section.
Remarkably, in this strong coupling regime, the eigenvalues
can be obtained explicitly and the infinite sum encountered in \eq{amp}
can be computed.

\subsection{Eigenvalues}
Let $d_i=de_i$,  
$\sum_{i=1}^3e_i^2=1$, and $d\gg 1$.  
For fixed $e_i$ and increasing $d$, the right-hand side 
of \eq{charac} increases in
magnitude, so the left-hand side must change accordingly. 
If the equality in \eq{charac} is positive, the 
eigenvalue $\l$ must move to the right (R). If it is negative, $\l$ must move
to the left (L).

Divide the real axis into six sections,
$J_1=(-\infty,m_1),\ J_2=(m_1,\l_b),\
J_3=(\l_b,m_2),\ J_4=(m_2,\l_c),\ J_5=(\l_c,m_3),\ J_6=(m_3,\infty)$, 
where $\l_b$ and $\l_c$ are the two zeros of 
\be
r(\l)\equiv\sum_{i=1}^3{e_i^2\over\l-m_i},\labels{r}\ee
explicitly given by
\be
\l_{b,c}&=&\h(p\mp c),\nn\\
p&=&e_1^2(m_2+m_3)+e_2^2(m_3+m_1)+e_3^2(m_1+m_2),\nn\\
c&=&\[p^2-4(e_1^2m_2m_3+e_2^2m_3m_1+e_3^2m_1m_2)\]^\h.\labels{lamab2}\ee
Then
$\l$ moves L if it is in $J_{1,3,5}$, and $\l$ moves R if it is in
$J_{2,4,6}$.  See Fig.~1.

Remember that if the real axis is divided into intervals $I_\l$ bounded by
the half integers and $m_i$, then
there is one and only one eigenvalue $\l$ in each interval. 
If $I_\l\subset J_{1,3,5}$, then $\l$ moves L.
If $I_\l\subset J_{2,4,6}$, then $\l$ moves R.
If $I_\l$ contains either $\l_b$ or $\l_c$, then $\l$ moves towards
these zeros of $r(\l)$.
In particular, if $m_i$ is one of the two boundaries of $I_\l$, then
$\l$ always moves away from that boundary. If both boundaries are
some $m_i$, then a zero $\l_{b,c}$ must be contained in that interval
where the eigenvalue moves towards. In any case, the eigenvalue always
stays away from the $m_i$ boundaries.

As $d\to\infty$, the eigenvalues will therefore end up either at the
half-integer boundaries, or a zero $\l_b$, or $\l_c$. 
In fact, there is one and only one eigenvalue at every half integer.

To see that, choose any half integer, and denote the interval to its
left (right) by $I_-$ ($I_+$). The only way this half integer ends
up not to be an eigenvalue, or end up to be equal to two different
eigenvalues, is when the two $\l$'s in $I_-$ and $I_+$ move
in opposite directions as $d$ increases. This however is not possible
unless a zero is contained in $I_-$ or in $I_+$. If it is in $I_-$, then
$\l\in I_-$ will drift towards this zero, and $\l\in I_+$ will move L,
drifting towards the half-integer value in question. Similar thing
happens when the zero is in $I_+$. We therefore conclude that there is
one and only eigenvalue at each half integer value.

It is important to note that the eigenvalues $\l_b$ and $\l_c$ are 
special simply because they do not fall into this 
infinite sequence
of half integers. It is not because they are brane states. 
In the strong coupling limit, brane and bulk components are thoroughly
mixed up. Nevertheless, we may still define a brane state as one that
extrapolates back to the eigenvalues $m_i$ when $d_i=0$, and a 
bulk state as one that extrapolates back to an integer eigenvalue
in that limit. With that definition,
each zero may correspond to either a brane state or a bulk state.
If $I_\l$ containing this zero  also contains an integer, 
then it is a bulk state. If it does not, then it must be a brane state.

\subsection{Transition amplitude}
With this strong-coupling spectrum, the transition amplitude in \eq{amp}
can be written more explicitly as
\be{\cal A}_{ij}(\t)&=&d^2e_ie_j\Bigg[\sum_{n=-\infty}^\infty 
\({e^{-i\l_n^2\tau}\over B^2(\l_n)(\l_n-m_i)(\l_n-m_j)}\)\nn\\
&&\hskip1.1cm +\ {e^{-i\l_b^2\tau}\over B^2(\l_b)(\l_b-m_i)(\l_b-m_j)}\nn\\
&&\hskip1.1cm+\ {e^{-i\l_c^2\tau}\over B^2(\l_c)(\l_c-m_i)
(\l_c-m_j)}\Bigg],
\labels{ampstrong}\ee

\bigskip
For $d\gg 1$, there is a qualitative difference between the normalization
factors $B^2(\l_n)$ and $B^2(\l_{b,c})$. This is so because the right-hand
side of \eq{charac}, which according to \eq{r} can be written as
$d^2r(\l)$, behaves differently for the two sets of eigenvalues. For
$\l_{b,c}$, $r(\l)=O(1/d^2)$ so these eigenvalues as well as
$\cos(\pi\l)$ are finite in the limit
$d\to\infty$. For $\l_n$, $r(\l)=O(d^2/(\l-m_i))$, so in the strong
coupling limit $\tan(\pi \l)$ as well as $1/cos(\pi\l)$ are of order
$d^2$ for finite $\l$, and are of order $O(d^2/\l)$ when $\l$ is 
comparable or larger than $d^2$. Using \eq{norm}, it follows then
that $B^2(\l)$ is of order $d^2$ for $\l_{b,c}$, but is of order
$d^4/\l$ for $\l_n$. 

Returning to \eq{ampstrong}, it follows from this observation 
that the second
and third terms on the right are finite in the strong coupling limit,
but each of the terms in the infinite sum is of order $1/d^2$ when
$\l_n\ll d^2$. This means that we can freely drop a finite number
(say $N$) of
terms from the infinite sum in the strong coupling limit, and 
retains only those terms in the infinite sum with large $\l_n$.
In that case, the right-hand side of \eq{charac} becomes $d^2/\l$,
and we can use it to solve for $1/\cos^2(\pi\l)$ to get 
$B^2(\l)\simeq 1/\cos^2(\pi\l)=1+d^2(1+\pi^2d^2)/\l^2\equiv 1+K^2/\lambda^2$. Since $\Delta\l_n\ll d^2$ and $N/d^2\ll 1$, 
we may now replace the
infinite sum on the right of \eq{ampstrong} by the integral
\be
f(\t)&=&\int_{-\infty}^\infty d\l{e^{-i\l^2\tau}\over \l^2+K^2}.\labels{f}\ee
At $\t=0$, the integral can be evaluated to give
$f(0)={\pi/K}\simeq {1/ d^2}$.

Moreover, for $d\gg 1$, it follows from \eq{norm} that
$B^2(\l)\simeq d^2s(\l)$
is true for $\l=\l_b,\l_c$ (assuming that $\cos(\l_{b,c})\not=0$), where
\be
s(\l)\equiv\sum_{i=1}^3\({e_i\over\l-m_i}\)^2.\labels{s}\ee
Therefore,
\be{\cal A}_{ij}(0)&=&e_ie_j\Bigg[
1+\ {1\over s(\l_b)(\l_b-m_i)(\l_b-m_j)}\nn\\
&&\hskip1.7cm+\ {1\over s(\l_c)(\l_c-m_i)(\l_c-m_j)}\Bigg].
\labels{unitarity}\ee
According to \eq{amp}, this should be equal to $\d_{ij}$ for all
parameters $e_i$ and $m_i$,
if the infinite-dimensional matrix
$U_{f\l}$ is unitary, as expected from the completeness relation.
This is indeed so and can be verified numerically. To prove 
the identity analytically, a change of parameters is desirable.

The model is specified by six parameters: $d_i=de_i$ and $m_i$.
They are actually not the most convenient parameters to use
because according to \eq{charac}, a shift of the
$m_i$'s by a common integer simply shifts all the eigenvalues by the
same integer, so the more useful parameters are $\l-m_i$ rather than
$m_i$. In fact, all the quantities in \eq{unitarity} can be expressed
in terms of the parameters
\be
x_i={1\over\l_b-m_i},\quad y_i={1\over\l_c-m_i}.\labels{xieta}\ee
Note that these two sets of parameters are related by
\be
{1\over y_i}-{1\over x_i}=\l_c-\l_b=c,\labels{constraint}\ee
so that there are only four independent parameters in the set: $c$ and $x_i$,
or $c$ and $y_i$.

We can use the three conditions, $e^2=1$ and $r(\l_b)=r(\l_c)=0$,
to solve for the three variables $e_1^2, e_2^2, e_3^2$. The result is
\be
e_1^2&=&(x_2 y_3-y_2 x_3)/D,\nn\\
e_2^2&=&( x_3 y_1- y_3 x_1)/D,\nn\\
e_3^2&=&( x_1 y_2- y_1 x_2)/D,\nn\\
D&=&( x_1 y_2+ x_2 y_3+ x_3 y_1)-
( y_1 x_2+ y_2 x_3+ y_3 x_1).\labels{e123}\ee
We can use \eq{constraint} to eliminate either $y_i$ or $ x_i$, and we get
respectively
\be
s_b\equiv s(\l_b)=c x_1 x_2 x_3,\labels{slb}\ee
or
\be
s_c\equiv s(\l_c)=-c y_1 y_2 y_3=-c{x_1x_2x_3\over(1+cx_1)(1+cx_2)(1+cx_3)}.
\labels{slc}\ee

In these notations, we can write \eq{unitarity} in a more transparent form as
\be
{\cal A}_{ij}(0)&=&\sum_{p=1}^3V^*_{ip}V_{jp},\labels{avv}\ee
where
\be
V&=&\pmatrix{e_1&e_1x_1/\sqrt{s_b}&e_1y_1/\sqrt{s_c}\cr
                  e_2&e_2x_2/\sqrt{s_b}&e_2y_2/\sqrt{s_c}\cr
                  e_3&e_3x_3/\sqrt{s_b}&e_3y_3/\sqrt{s_c}},\labels{v}\ee
 up to an arbitrary phase for each column. 
The matrix $V$ is unitary, as can be verified from
the explicit equations \eq{constraint} to \eq{slc}.

The matrix $V$ is parametrized  by four real parameters, $c,x_i$,
plus three arbitrary phases of $e_i$, more than the usual $3 \times 3$
mixing matrix. 
Nevertheless, there are mixing matrices not of the form \eq{v},
with the constraint \eq{constraint} to \eq{slc}.
An example is the bimaximal neutrino mixing matrix ($s_3=\sin\theta_3$, $
c_3=\cos \theta_3$):
\be
u=\pmatrix{{c_3/\sqrt{2}}&{c_3/\sqrt{2}}&s_3\cr 
-\h(1+s_3)&\h(1-s_3)&{c_3/\sqrt{2}}\cr
\h(1-s_3)&-\h(1+s_3)&{c_3/\sqrt{2}}\cr}.\labels{bimaximal}\ee

We are now ready to discuss the case
$\t>0$. The function $f(\t)$ can be expressed in terms of
error function with an imaginary argument, but it is more straight forward
to write it directly in the form
$f(\t)={(\pi/ K)}g(K^2\tau)$, where
\be
g(x)&\equiv&{1\over\pi}\int_{-\infty}^\infty 
d u{e^{-i u^2x}\over u^2+1}.\labels{g}\ee
 The absolute magnitude of the function $g(x)$ decreases monotonically
from 1 to zero. For large $x$, $g(x)\simeq (1-i)/\sqrt{2\pi x}$.

The transition amplitude (19), in the strong coupling limit $d\gg 1$, is 
therefore equal to
\be{\cal A}_{ij}(\t)&=&e_ie_j\Bigg[g(K^2\t)
+\ {e^{-i\l_b^2\tau}\over s(\l_b)(\l_b-m_i)(\l_b-m_j)}\nn\\
&&\hskip2.4cm+\ {e^{-i\l_c^2\tau}\over s(\l_c)(\l_c-m_i)(\l_c-m_j)}\Bigg]
\nn\\
&=&V^*_{i1}V_{j1}g(K^2\t)+V^*_{i2}V_{j2}e^{-i\l_b^2\tau}
+V^*_{i3}V_{j3}e^{-i\l_c^2\tau},
\labels{ampfin}\ee
with the matrix $V$ given in \eq{v}.
At any `time' $\tau$, the amplitude is described by six parameters:
$K,c,x_1,x_2,x_3$, and one of the $m_i$'s so that $\l_b$
and $\l_c$ can be determined from $x_i$ and $y_i$ respectively.
Alternatively, we can let $\l_b$ be the sixth parameter, whence
$\l_c=\l_b+c$.

\subsection{Comparison with direct mixing}
In this theory, neutrino mixing is
induced by coupling of active
neutrinos to bulk neutrinos. No direct mixing of 
active neutrinos takes place. Nevertheless, in the strong coupling
limit, the resulting amplitude
\eq{ampfin} bears a close resemblance to the usual theory in which
sterile neutrino is absent, and mixing of active neutrinos is
directly introduced through the $3\x3$ MNS matrix $u$ \cite{MNS}.
If $\mu_1,\mu_2,\mu_3$ are the mass eigenstates, the transition
amplitude with direct mixing is given by 
\be{\cal A}_{ij}(\t)&=&
u^*_{i1}u_{j1}e^{-i\mu_1^2\t}+u^*_{i2}u_{j2}e^{-i\mu_2^2\tau}
+u^*_{i3}u_{j3}e^{-i\mu_3^2\tau}.
\labels{ampmns}\ee
Since both $V$ and $u$ are unitary matrices, if we identify
$\l_b^2$ with $\mu_2^2$ and $\l_c^2$ with $\mu_3^2$ (modulo a common
constant), then the main difference between
\eq{ampfin} and \eq{ampmns} is the replacement of $g(K^2\t)$ by
$e^{-i\mu_1^2\t}$. Both functions equal to unity at $\t=0$, but the 
latter oscillates and the former decays to zero within
a short transient `time' $\sim 1/K^2\ll 1$.
The oscillations mediated by three mass eigenstates in the direct-mixing
theory, is replaced after the transient period by 
an oscillation mediated by two eigenstates
($\l_b$ and $\l_c$) in the bulk theory.
Note that these two eigenstates can be either brane states
or bulk states (see the discussion at the end of Sec.~IIIA), so in
general they have no direct connection with the three mass eigenstates
in the direct-mixing theory. In particular, we should not think that
one of these three mass eigenstates gets lost after the transient period.

Moreover, unlike the direct-mixing theory in which the 
total probability to be in some active neutrino state is always  1, 
in the bulk theory this total probability decreases with $\t$
because of leakages into the bulk. Quantitatively,
\be
{\cal P}_j(\t)\equiv\sum_{i=1}^3{\cal P}_{ij}(\t)=1-e_j^2\{1-
|g(K^2\t)|^2\}\simeq 1-e_j^2\quad(\t\gg K^{-2}).\labels{totprobend}\ee

 As a check we go to the limit of equal couplings, i.e. all $e_i$'s are equal. 
Since
$\sum_i e_i^2 = 1$ then each $e_i^2 =1/3$ and the above 
probability approached 2/3 in agreement
with the result of \cite{KEITH}.

In the bulk neutrino theory, with an infinite number of sterile
neutrinos, is very different from the direct-mixing theory
with three active
neutrinos, or three active neutrino plus one sterile neutrino. 
In particular, there is no way
the bulk neutrino theory can explain both the solar and the atmospheric
neutrino deficits simultaneously, at least not in the strong-coupling limit 
after the transient period is passed.

To see that, we use \eq{ampfin} to write the survival probability after
the transient period in the form
\be
{\cal P}_{ii}(\t)&=&|{\cal A}_{ii}(\t)|^2=p_{i}-p'_{i}\cos^2\(\h\D^2\t\),
\nn\\
p_{i}&=&\(|V_{i2}|^2+|V_{i3}|\)^2=\(1-e_i\)^2,\nn\\
p'_{i}&=&4(1-e_i^2)|V_{i3}|^2,\nn\\
\D^2&=&\l_c^2-\l_b^2.\labels{probfin}\ee
This is to be compared with the usual two-term survival probability formula
for solar and atmospheric neutrinos,
\be
{\cal P}_i(\t)&=&1-\sin^2(2\theta_i)\sin^2\(\h\D_i^2\t\),
\labels{probmns}\ee
where $\D_1^2$ for solar neutrino oscillations is at least two orders of
magnitude smaller than $\D_2^2$ for atmospheric neutrino oscillations.
There is no way the formula \eq{probfin}, with $\D$ independent of $i$,
can fit both the solar and the atmospheric deficits.

We might want to be less ambitious and use \eq{probfin} to fit only
solar (or atmospheric, but not both) neutrino deficit by adjusting
the parameters for \eq{probfin} to emulate \eq{probmns} with
the experimentally measured $\theta_i$ and $\D_i=\D$. Even so it will
not work, not exactly anyway. The reason is that to make $p_i=1$, 
we must put $e_i=0$. In that case \eq{r} has only one but not two zeros,
and the three terms appearing in \eq{ampfin} become two terms. After
the transient period, we are left with one term, whose probability
does not oscillate at all.

These failures are due to the disappearance of the first term beyond
the transient period. What about taking the opposite limit when
$\t$ is so small that $g(K^2\t)\simeq 1$? That would not work either
for two reasons. First, in order for oscillations to occur we must
have $\l_b^2\t$ and/or $\l_c^2\t$ to be of order 1, which requires
$\l_b^2\gg K^2$ and/or $\l_c^2\gg K^2$. This cannot be achieved
within the strong coupling limit when $e_i$ and $m_i$ are fixed
and $d\gg 1$. Moreover, even if we ignore this difficulty, there is
no way the unitary matrix $V$ can simulate a realistic MNS matrix $u$,
as pointed out previously in \eq{bimaximal}.

In conclusion, the idea of using a bulk neutrino living in an extra dimension
to explain neutrino oscillation is a very attractive one. It might explain
why neutrino masses are small and in the strong brane-bulk
coupling limit, why their mixings are big. Unfortunately, such strong
coupling also causes a large leakage into the sterile modes in the bulk,
so that this theory in the strong-coupling limit
is not compatible with experiments.
Whether an intermediate coupling is viable or not depends
on whether rapid variations as a function of energy and distance are
observed in future experiments. 

Our study complements the mostly numerical work done on this simple 
model. Since the solution is nonperturbative, we deem 
this work to be an important first in the study of
the physics of strongly coupled bulk neutrinos. As more data in 
solar and atmospheric neutrinos
become available there will be more serious attempts at seeing 
whether bulk neutrinos can
explain all or part of such improved data sets. 
Our work can be used as a guide to such
attempts.

This research is supported by the Natural Science and Engineering
Research Council of Canada, and the research of CSL is also supported
by the  Qu\'ebec Department of Education. This investigation was initiated
during a visit of CSL to TRIUMF. He wishes to thank the Theory Group
there for its hospitality. He also wishes to thank
Keith Dienes for informative discussions. JNN would also like to thank 
Professor T.K.Lee of the National Center for Theoretical Science, Taiwan
for his hospitality where part of this research is done.

\newpage

\begin{figure}[h]
\includegraphics{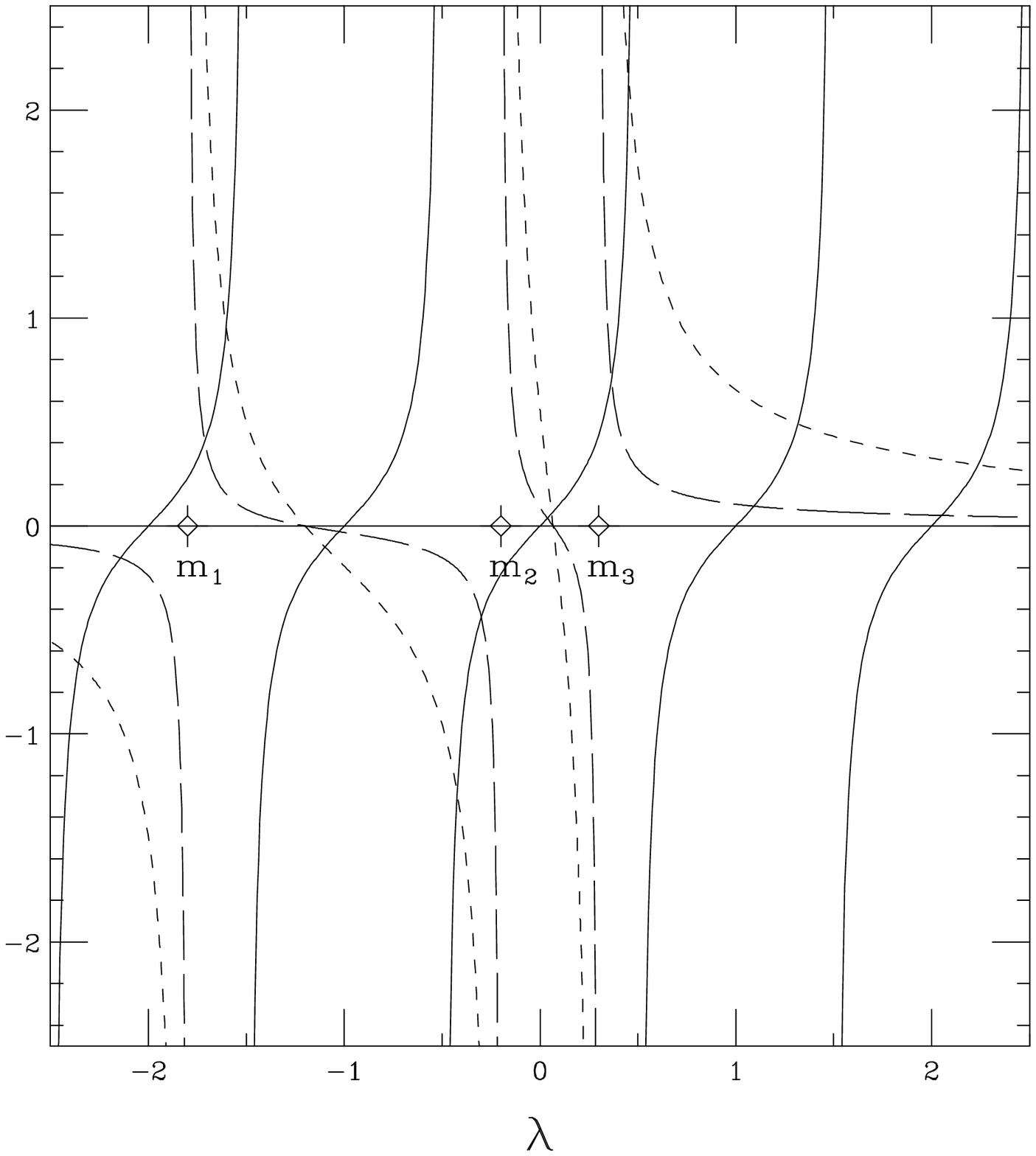}
\vspace{15cm}
\caption{Graphical solutions of the characteristic equation \eq{charac}.
The left hand side of \eq{charac} is represented by the solid curve; the
right hand side is represented by the dashed curves, with $(m_1,m_2,m_3)=
(-1.8,-0.2,0.3)$. The one with the short dash has the larger coupling
constant $d$ than the one with the long dash. Note that the solution
$\l$ moves up and right along the solid curve for increasing $d$
when the value on both
sides of \eq{charac} is positive, and moves down and left along the solid
curve when the value is negative.}
\end{figure}


\begin{thebibliography}{9}
\bibitem{KK} T. Kaluza, Math.~Phys.~Kl. (1921) 966; O. Klein, Phys.~Z. 37 
(1926) 895.
\bibitem{POL} J. Polchinski, Phys.~Rev.~Lett. 75 (1995) 4724.
\bibitem{AK} N. Arkani-Hamed, S. Dimopoulos, and G. Dvali,
Phys.~Lett. B429 (1998) 263; Phys.~Rev. D59 (1999) 086004;
I. Antoniadis, N. Arkani-Hamed, S. Dimopoulos, and G. Dvali,
Phys.~Lett. B436 (1998) 257; L. Randall and R. Sundrum, Phys.~Rev.~Lett. 
83(1999) 3370, 4690. 
\bibitem{WASH} J.C. Long, A.B. Churnside, and J.C. Price;
C.D. Hoyle et. al., Phys.~Rev.~Lett. 86 (2001) 1418.
\bibitem{COS}
See, {\it e.g.}, J.M. Cline, C. Grojean and G. Servant,
Phys.~Rev.~Lett. 83(1999) 4245;
P. Binetruy, C. Deffayet, U. Ellwanger and D. Langlois,
Phys.~Lett.  B477 (2000) 285;
C. Csaki, M. Graesser, L. Randall and J. Terning,
Phys.~Rev. D62 (2000) 045015, and references therein.

\bibitem{KEITH}  K.R. Dienes, E. Dudas, and T. Gherghetta,
Nucl.~Phys. B557 (1999) 25; K.R. Dienes and
I. Sarcevic, Phys.~Lett. B500 (2001) 133-141;
D.O. Caldwell, R.N. Mohapatra, and S.J. Yellin,
hep-ph/0010353; 0102279.

\bibitem{MO} N. Arkani-Hamed, S. Dimopoulos, G. Dvali, and J. March-Russell,
hep-ph/9811448;
G. Dvali and A.Yu. Smirnov, Nucl.~Phys. B563 (1999) 63;
R.N. Mohapatra, S. Nandi, and A. P\'erez-Lorenzana, Phys.~Lett. B466 (1999) 
115; Nucl.~Phy. B593 (2001);
R. Barbieri, P. Creminelli, and A. Strumia,
Nucl.~Phys. B585 (2000) 28; 
R.N. Mohapatra and A. P\'erez-Lorenzana, Nucl.~Phys.B 576 
(2000) 466; Nucl.~Phys. B593 (2001) 451;
N. Cosme, J.-M. Fr\`ere, Y. Gouverneur, F.-S.Ling, D. Monderen, and V.
Van Elewyck, Phys.~Rev. D63 (2001) 113018.

\bibitem{NEU}  A.E. Faraggi and M. Pospelov, Phys.~Lett. B458 (1999) 237;
A. Das and O.C.W. Kong, Phys.~Lett. B470 (1999) 149; 
G.C. McLaughlin and J.N. Ng, Phys.~Lett. B470 (1999) 157;
B493 (2000) 88;
A. Lukas, P. Ramond, A. Romanino, and G.G. Ross, 
Phys.~Lett. B495 (2000) 136; hep-ph/0011295. 

\bibitem{BAHCALL} J.N. Bahcall, P.I. Krastev, and A.Yu. Smirnov,
hep-ph/0103179.
\bibitem{MAINZ} J. Bonn {\it et al}., Nucl.~Phys.~B (Proc. Suppl.) 
91 (2001) 273.
\bibitem{HM} H.V. Klapdor-Kleingrothaus, hep-ph/0102319.
\bibitem{SK} The Super-Kamiokande Collaboration,
Phys.~Rev.~Lett. 85 (2000) 3999-4003 (hep-ex/0009001); 
H. Sobel, Nucl.~Phys.~B (Proc. Suppl.) 91 (2001) 127;
Y. Suzuki, Nucl.~Phys.~B (Proc. Suppl.) 91 (2001) 29.

\bibitem{MARRONE} G.I. Fogli, E. Lisi, and A. Marrone, 
Phys.~Rev. D63 (2001) 053008.

\bibitem{MNS} Z. Maki, M. Nakagawa, and S. Sakata, Prog.~Theo.~Phys. 28
(1962) 870.
\end{thebibliography}
\end{document}